\documentclass[conference]{IEEEtran}
\usepackage{xcolor}
\usepackage{hyperref}
\usepackage[pdftex]{graphicx}
\usepackage{caption}

\usepackage[ruled,linesnumbered]{algorithm2e}
\usepackage{algpseudocode} 
\usepackage{multirow}
\usepackage{listings}
\usepackage{booktabs} 

\usepackage{array}    
\newcolumntype{L}[1]{>{\raggedright\let\newline\\\arraybackslash\hspace{0pt}}m{#1}}
\newcolumntype{C}[1]{>{\centering\let\newline\\\arraybackslash\hspace{0pt}}m{#1}}
\newcolumntype{R}[1]{>{\raggedleft\let\newline\\\arraybackslash\hspace{0pt}}m{#1}}

\begin{document}
\title{HW2VEC: A Graph Learning Tool for Automating Hardware Security}
\author{
    \IEEEauthorblockN{
        Shih-Yuan Yu\textsuperscript{\textsection}, 
        Rozhin Yasaei\textsuperscript{\textsection}, 
        Qingrong Zhou,
        Tommy Nguyen,
        Mohammad Abdullah Al Faruque
    }
    \IEEEauthorblockA{
        \textit{Department of Electrical Engineering and Computer Science} \\
        \textit{University of California, Irvine, California, USA}\\ 
        \textit{\{shihyuay, ryasaei, qingronz, tommytn1, alfaruqu}@uci.edu\}
    }
}

\maketitle

\begingroup\renewcommand\thefootnote{\textsection}
\footnotetext{Both are equal-contributing first authors. Yu is the corresponding author.}
\endgroup

\begin{abstract}
The time-to-market pressure and continuous growing complexity of hardware designs have promoted the globalization of the Integrated Circuit (IC) supply chain.
However, such globalization also poses various security threats in each phase of the IC supply chain.
Although the advancements of Machine Learning (ML) have pushed the frontier of hardware security, most conventional ML-based methods can only achieve the desired performance by manually finding a robust feature representation for circuits that are non-Euclidean data.
As a result, modeling these circuits using graph learning to improve design flows has attracted research attention in the Electronic Design Automation (EDA) field.
However, due to the lack of supporting tools, only a few existing works apply graph learning to resolve hardware security issues.
To attract more attention, we propose \textsc{HW2VEC}, an open-source graph learning tool that lowers the threshold for newcomers to research hardware security applications with graphs.
\textsc{HW2VEC} provides an automated pipeline for extracting a graph representation from a hardware design in various abstraction levels (register transfer level or gate-level netlist). 
Besides, \textsc{HW2VEC} users can automatically transform the non-Euclidean hardware designs into Euclidean graph embeddings for solving their problems.
In this paper, we demonstrate that \textsc{HW2VEC} can achieve state-of-the-art performance on two hardware security-related tasks: \textit{Hardware Trojan Detection} and \textit{Intellectual Property Piracy Detection}.
We provide the time profiling results for the graph extraction and the learning pipelines in \textsc{HW2VEC}.
\end{abstract}

\IEEEpeerreviewmaketitle

\section{Introduction}
In past decades, the growing design complexity and the time-to-market pressure have jointly contributed to the globalization of the Integrated Circuit (IC) supply chain~\cite{shamsi2019ip}.
Along this globalized supply chain, IC designers tend to leverage third-party Electronic Design Automation (EDA) tools and Intellectual Property (IP) cores or outsource costly services to reduce their overall expense.
This results in a worldwide distribution of IC design, fabrication, assembly, deployment, and testing~\cite{board2005defense, jose2008innovation, rostami2014primer}.
However, such globalization can also make the IC supply chain vulnerable to various hardware security threats such as \textit{Hardware Trojan Insertion}, \textit{IP Theft}, \textit{Overbuilding}, \textit{Counterfeiting}, \textit{Reverse Engineering}, and \textit{Covert \& Side Channel Attacks}. 

As the consequences of not promptly addressing these security threats can be severe, countermeasures and tools have been proposed to mitigate, prevent, or detect these threats~\cite{hu2020overview}. 
For example, hardware-based primitives, physical unclonable functions (PUFs)~\cite{herder2014physical}, true random number generator (TRNG)~\cite{rahman2014ti}, and cryptographic hardware can all intrinsically enhance architectural security.
The countermeasures built into hardware design tools are also critical for securing the hardware in the early phases of the IC supply chain. 
Some Machine Learning (ML) based approaches have been proven effective for detecting \textit{Hardware Trojans} (HT) from hardware designs in both Register Transfer Level (RTL) and Gate-Level Netlist (GLN)~\cite{han2019hardware, hasegawa2018hardware}.
Besides,~\cite{huang2012parametric} automates the process of identifying the counterfeited ICs by leveraging Support Vector Machine (SVM) to analyze the sensor readings from on-chip hardware performance counters (HPCs). 
However, as indicated in~\cite{tan2020challenges}, effectively applying ML models is a non-trivial task as the defenders must first identify an appropriate input representation based on hardware domain knowledge.
Therefore, ML-based approaches can only achieve the desired performance with a robust feature representation of a circuit (non-Euclidean data) which is more challenging to acquire than finding the one for Euclidean data such as images, texts, or signals.

\begin{figure}[!h]
    \centering
    \includegraphics[width=1.0\linewidth]{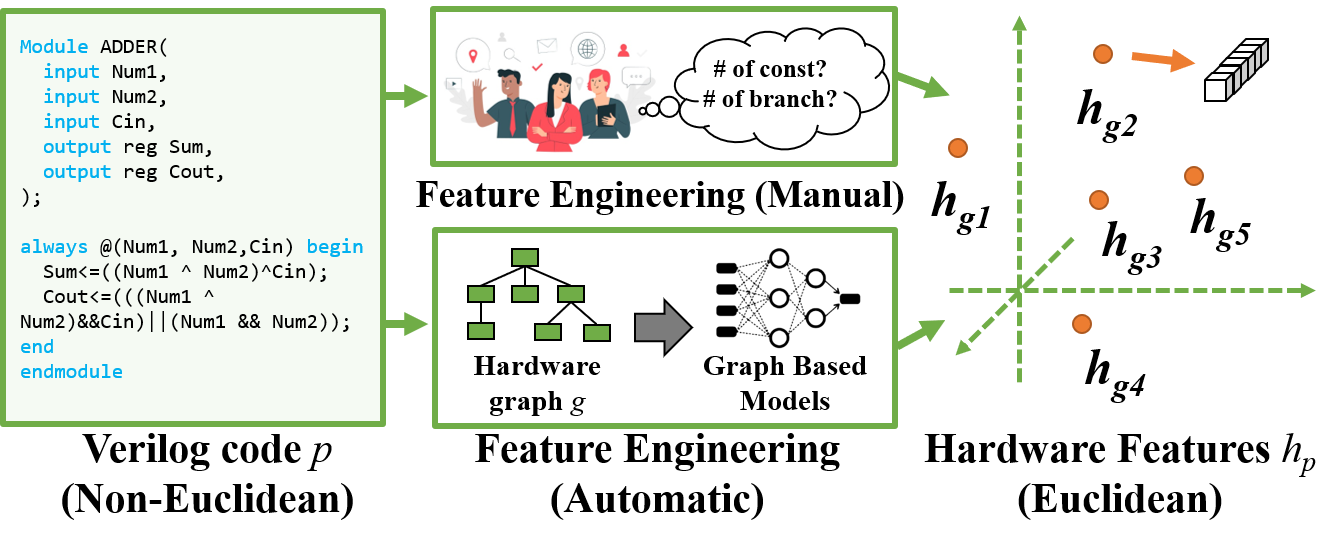}
    \caption{The illustration of the process that extracts features for hardware analysis.}
    \label{fig:graphex}
\end{figure}

In IC design flow, many fundamental objects such as netlists or layouts are natural graph representations~\cite{ma2020understanding}.
These graphs are non-Euclidean data with irregular structures, thus making it hard to generalize basic mathematical operations and apply them to conventional Deep Learning (DL) approaches~\cite{cai2018comprehensive}. 
Also, extracting a feature that captures structural information requires a non-trivial effort to achieve the desired performance.
To overcome these challenges, many \textit{graph learning} approaches such as \textit{Graph Convolutional Networks} (GCN), \textit{Graph Neural Networks} (GNN), or \textit{Graph Autoencoder} (GAE) have been proposed and applied in various applications such as computer vision, natural language processing, and program analysis~\cite{kipf2016semi, wu2020comprehensive}.
In the EDA field, some works tackle netlists with GCNs for test point insertion~\cite{ma2019high} or with GNNs for fast and accurate power estimation in pre-silicon simulation~\cite{zhang2020grannite}.
As Figure~\ref{fig:graphex} shows, these approaches typically begin with extracting the graph representation ($g$) from a hardware design $p$, then use the graph-based models as an alternative to the manual feature engineering process. 
Lastly, by projecting each hardware design onto the Euclidean space ($h_g$), these designs can be passed to ML models for learning tasks.
However, only a few works have applied GNN-based approaches for securing hardware during IC design phases due to the lack of supporting tools~\cite{gnn4ip2021, gnn4tj2021}.

To attract more research attention to this field, we propose \textsc{HW2VEC}, an open-source graph learning tool for enhancing hardware security.
\textsc{HW2VEC} provides automated pipelines for extracting graph representations from hardware designs and leveraging graph learning to secure hardware in design phases.
Besides, \textsc{HW2VEC} automates the processes of engineering features and modeling hardware designs.  
To the best of our knowledge, \textsc{HW2VEC} is the first open-source research tool that supports applying graph learning methods to hardware designs in different abstraction levels for hardware security. 
In addition, \textsc{HW2VEC} supports transforming hardware designs into various graph representations such as the \textit{Data-Flow Graph} (DFG), or the \textit{Abstract Syntax Tree} (AST). 
In this paper, we also demonstrate that \textsc{HW2VEC} can be utilized in resolving two hardware security applications: \textit{Hardware Trojan Detection} and \textit{IP Piracy Detection} and can perform as good as the state-of-the-art GNN-based approaches.

\subsection{Our Novel Contributions}
Our contributions to the hardware security research community are as follows,
\begin{itemize}
    \item We propose an automated pipeline to convert a hardware design in RTL or GLN into various graph representations. 
    \item We propose a GNN-based tool to generate vectorized embeddings that capture the behavioral features of hardware designs from their graph representations.
    \item We demonstrate \textsc{HW2VEC}'s effectiveness by showing that it can perform similarly compared to state-of-the-art GNN-based approaches for various real-world hardware security problems, including \textit{Hardware Trojan Detection} and \textit{IP Piracy Detection}.
    \item We open-source \textsc{HW2VEC} as a Python library\footnote{The  \textsc{HW2VEC} is publicly available at \url{https://github.com/AICPS/hw2vec/}.} to contribute to the hardware security research community.
\end{itemize}

\subsection{Paper Organization}
We organize the rest of the paper as follows: we introduce background information and literature survey in Section~\ref{sec:related_works}; we present the overall architecture of \textsc{HW2VEC} in Section~\ref{sec:hw2vec}; then, we demonstrate the usage examples and two advanced use-cases (HT detection and IP piracy detection) in Section~\ref{sec:use-case}; Next, we show experimental results and discuss \textsc{HW2VEC}'s practicability in Section~\ref{sec:exp}; Lastly, we conclude in Section~\ref{sec:conclusion}.
\section{Related works and Background}
\label{sec:related_works}
This section first briefly overviews hardware security problems and countermeasures. 
Then it describes the works applying ML-based approaches for hardware security. Lastly, we introduce the works that utilize graph learning methods in both EDA and hardware security.

\subsection{Hardware Security Threats in IC Supply Chain}
\begin{figure*}[!ht]
    \centering
    \includegraphics[width=1.0\linewidth]{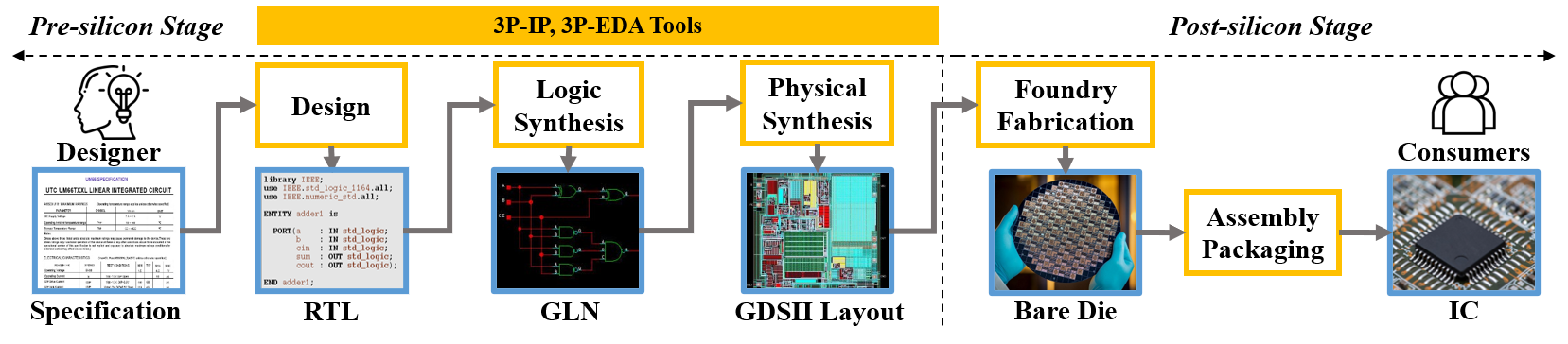}
    \caption{The illustration of the IC supply chain demonstrating the hardware design flow from a specification to the behavioral description (RTL), logic implementation (GLN), physical implementation (GDSII), and the actual chip (Bare Die or IC).}
    \label{fig:chain}
\end{figure*}

In the IC supply chain, each IC is passed through multiple processes as shown in Figure~\ref{fig:chain}.
First, the specification of a hardware design is turned into a behavioral description written in a Hardware Design Language (HDL) such as Verilog or VHDL.
Then, it is transformed into a design implementation in terms of logic gates (i.e., netlist) with \textit{Logic Synthesis}.
\textit{Physical Synthesis} implements the netlist as a layout design (e.g., a GDSII file).
Lastly, the resulting GDSII file is handed to a foundry to fabricate the actual IC. 
Once a foundry produces the IC (Bare Die), several tests are performed to guarantee its correct behavior.
The verified IC is then packaged by the assembly and sent to the market to be deployed in systems.

For a System-on-Chip (SoC) company, all of the mentioned stages of the IC supply chain require a vast investment of money and effort.
For example, it costs \$5 billion in 2015 to develop a new foundry~\cite{yeh2012trends}.
Therefore, to lower R\&D cost and catch up with the competitive development cycle, an SoC company may choose to outsource the fabrication to a third-party foundry, purchase third-party IP cores, and use third-party EDA tools.
The use of worldwide distributed third parties makes the IC supply chain susceptible to various security threats~\cite{xiao2016hardware} such as \textit{Hardware Trojan Insertion}, \textit{IP Theft}, \textit{Overbuilding}, \textit{Counterfeiting}, \textit{Reverse Engineering}, and \textit{Covert \& Side Channel Attacks}, etc.
Not detecting or preventing these threats can lead to severe outcomes. 
For example, in 2008, a suspected nuclear installation in Syria was bombed by Israeli jets because a backdoor in its commercial off-the-shelf microprocessors disabled Syrian radar~\cite{syrianRadar}.
In another instance, the IP-intensive industries of the USA lose between \$225 to \$600 billion annually as the companies from China steal American IPs, mainly in the semiconductor industry~\cite{news2}.

Among the mentioned security threats, the insertion of \textit{Hardware Trojan} (HT) can cause the infected hardware to leak sensitive information, degrade its performance, or even trigger a Denial-of-Service (DoS) attack. 
In System-on-Chip (SoC) or IC designs, \textit{IP Theft}, the illegal usage and distribution of an IP core can occur. 
The third-party foundries responsible for outsourced fabrication can \textit{overbuild} extra chips just for their benefits without the designer's permission.
Moreover, selling the \textit{Counterfeited} designs in the name of its original supplier leads to financial or safety damage to its producer or even the national security if the target is within essential infrastructures or military systems.
\textit{Reverse engineering} (RE) recovers the high-level information from a circuit available in its lower-level abstraction. Although RE can be helpful in the design and verification process, an attacker can misuse the reconstructed IC designs for malicious intentions. 
\textit{Covert Channel} uses non-traditional communication (e.g., shared cache) to leak critical information of a circuit. 
In contrast, \textit{Side Channel} exists among the hardware components that are physically isolated or not even in proximity (e.g., power or electromagnetic channel).

\subsection{Hardware Security Countermeasures}

Due to the globalization of the IC supply chain, the hardware is susceptible to security threats such as IP piracy (unlicensed usage of IP), overbuilding (unauthorized manufacturing of the circuit), counterfeiting (producing a faithful copy of circuit), reverse engineering, hardware Trojan (malicious modification of circuit), and side-channel attacks \cite{ashraf2018towards}. 

In the literature, countermeasures and tools have been proposed to mitigate, prevent, or detect these threats~\cite{hu2020overview}.
For example, a cryptographic accelerator is a hardware-based countermeasure that can reinforce the build-in instead of the add-on defense against security threats.
True Random Number Generator (TRNG) and Physical Unclonable Function (PUF) are two other effective security primitives~\cite{herder2014physical, rahman2014ti}. 
These solutions are critical for security protocols and unique IC identification, and they rely on the physical phenomena for randomness, stability, and uniqueness, such as process variations during fabrication~\cite{tan2020challenges}.

In addition to hardware-based solutions, countermeasures enhancing the security during the hardware design process are also present in the literature.
For example, side-channel analysis for HT detection using various models such as hierarchical temporal memory \cite{htm} and DL~\cite{htnet} has grabbed lots of attention recently. However, they postpone the detection to \textit{post-silicon} stage.
On the other hand, \textit{Formal Verification} (FV) is a \textit{pre-silicon} algorithmic method which converts the 3PIP to a proof checking format and checks if the IP satisfies some predefined security properties~\cite{cadense2013jasper, subramanyan2014formal}.
Although FV leverages the predefined security properties in IP for HT detection, its detection scope is limited to certain types of HTs because the properties are not comprehensive enough to cover all kinds of malicious behaviors~\cite{rajendran2015detecting}. 
Some works employ model checking but are not scalable to large designs as model checking is NP-complete and can suffer from state explosion~\cite{rajendran2016formal}.
Another existing approach is \textit{code coverage} which analyzes the RTL code using metrics such as line, statement, finite state machine, and toggle coverage to ascertain the suspicious signals that imitate the HT~\cite{waksman2013fanci, zhang2015veritrust}.

As for IP theft prevention, \textit{watermarking} and \textit{fingerprinting} are two approaches that embed the IP owner and legal IP user's signatures into a circuit to prevent infringement~\cite{poudel2020flashmark, rai2019hardware}. 
\textit{Hardware metering} is an IP protection method in which the designer assigns a unique tag to each chip for chip identification (passive tag) or enabling/disabling the chip (active tag)~\cite{koushanfar2017active}.
\textit{Obfuscation} is another countermeasure for IP theft~\cite{chen2020decoy} which comprises two main approach; \textit{Logic Locking} and \textit{Camouflaging}.
In \textit{Logic Locking}, the designer inserts additional gates such as XOR into non-critical wires. 
The circuit will only be functional if the correct key is presented in a secure memory out of reach of the attacker~\cite{xie2017delay}. 
\textit{Camouflaging} modifies the design such that cells with different functionalities look similar to the attacker and confuses the reverse engineering process~\cite{camouflaging}. 
Lastly, another countermeasure is to split the design into separate ICs and have them fabricated in different foundries so that none of them has access to the whole design to perform malicious activities~\cite{patnaik2018raise, zhang2018analysis}.

In~\cite{hu2020overview}, several academic and commercial tools have been proposed to secure hardware. 
For example, \textit{VeriSketch}, \textit{SecVerilog}, etc., are the open-source academia verification tools for securing hardware.
\textit{SecureCheck} from \textit{Mentor Graphics}, \textit{JasperGold Formal Verification Platform} from \textit{Cadence}, and \textit{Prospect} from \textit{Tortuga Logic} are all commercial verification tools ready in the market.
\textit{PyVerilog}~\cite{Takamaeda:2015:ARC:Pyverilog} is a hardware design tool that allows users to parse HDL code and perform \textit{pre-silicon} formal verification side-by-side with functional verification.
In short, though many approaches have been proposed to counteract security threats, security is still an afterthought in hardware design. Therefore, new countermeasures will be needed against new security threats.

\subsection{Machine Learning for Hardware Security}
In the last few decades, the advancements in Machine Learning (ML) have revolutionized the conventional methods and models in numerous applications throughout the design flow. 
Defenders can use ML with hardware-based observations for detecting attacks, while attackers can also use ML to steal sensitive information from an IC, breaching hardware security~\cite{tan2020challenges}.
Some ML-based countermeasures have been proven effective for detecting HT from hardware designs in both Register Transfer Level (RTL) or gate-level netlists (GLN)~\cite{han2019hardware, hasegawa2018hardware}.
In~\cite{han2019hardware}, the circuit features are extracted from the Abstract Syntax Tree (AST) representations of RTL codes and fed to gradient boosting algorithm to train the ML model to construct an HT library.
\cite{hasegawa2018hardware} extracts 11 Trojan-net feature values from GLNs and then trains a Multi-Layer Neural Network on them to classify each net in a netlist as a normal netlist or Trojan.
Similarly, researchers have applied ML for automating the process of detecting other threats. 
For instance, SVM can be used to analyze the on-chip sensor readings (e.g., HPCs) to identify counterfeited ICs and detect HT in real-time~\cite{huang2012parametric, kulkarni2016svm}.
However, as indicated in~\cite{tan2020challenges}, effectively applying ML models is not a trivial task as the defenders must first identify an appropriate input representation for a hardware design.
Unlike Euclidean data such as images, texts, or signals, finding a robust feature representation for a circuit (Non-Euclidean data) is more challenging as it requires domain knowledge in both hardware and ML.
To overcome this challenge, \textsc{HW2VEC} provides more effective graph learning methods to automatically find a robust feature representation for a non-Euclidean hardware design.

\subsection{Graph Learning for Hardware Design and Security}
Although conventional ML and DL approaches can effectively capture the features hidden in Euclidean data, such as images, text, or videos, there are still various applications where the data is graph-structured.
As graphs can be irregular, a graph can have a variable size of unordered nodes, and nodes can have a different number of neighbors, thus making mathematical operations used in deep learning (e.g., 2D Convolution) challenging to be applied~\cite{cai2018comprehensive}.
Also, extracting a feature that captures structural information requires challenging efforts to achieve the desired performance.
To address these challenges, recently, many \textit{graph learning} approaches such as \textit{Graph Convolutional Networks} (GCN), \textit{Graph Neural Networks} (GNN), or \textit{Graph Autoencoder} (GAE) have been proposed and applied in various applications~\cite{kipf2016semi, wu2020comprehensive}.
Only by projecting non-Euclidean data into low-dimensional embedding space can the operations in ML methods be applied.

In EDA applications, many fundamental objects such as Boolean functions, netlists, or layouts are natural graph representations~\cite{ma2020understanding}.
Some works tackle netlists with GCNs for test point insertion~\cite{ma2019high} or with GNNs for fast and accurate power estimation in \textit{pre-silicon} simulation~\cite{zhang2020grannite}.
\cite{zhang2020grannite} uses a GNN-based model to infer the toggle rate of each logic gate from a netlist graph for fast and accurate average power estimation without gate-level simulations, which is a slower way to acquire toggle rates compared to RTL simulation.
They use GLNs, corresponding input port, and register toggle rates as input features and logic gate toggle rates as ground-truth to train the model. 
The model can infer the toggle rate of a logic gate from input features acquired from RTL simulation for average power analysis computed by other power analysis tools.

As for hardware security, only a few works utilizing GNN-based approaches against security threats exist~\cite{gnn4ip2021, gnn4tj2021}. 
\cite{gnn4tj2021} utilizes a GNN-based approach for detecting HT in \textit{pre-silicon} design phases without the need for golden HT-free reference. Besides, using the GNN-based approach allows the extraction of features from Data-Flow graphs to be automated.
In \cite{gnn4ip2021}, the proposed GNN-based approach can detect IP piracy without the need to extract hardware overhead to insert signatures to prove ownership. 
Specifically, the Siamese-based network architecture allows their approach to capturing the features to assess the similarity between hardware designs in the form of a Data-Flow Graph.
In short, these works have shown the effectiveness of securing hardware designs with graph learning approaches.
To further attract attention, we propose \textsc{HW2VEC} as a convenient research tool that lowers the threshold for newcomers to make research progress and for experienced researchers to explore this topic more in-depth.
\section{HW2VEC Architecture}
\label{sec:hw2vec}
\begin{figure*}[!ht]
    \centering
    \includegraphics[width=1.0\linewidth]{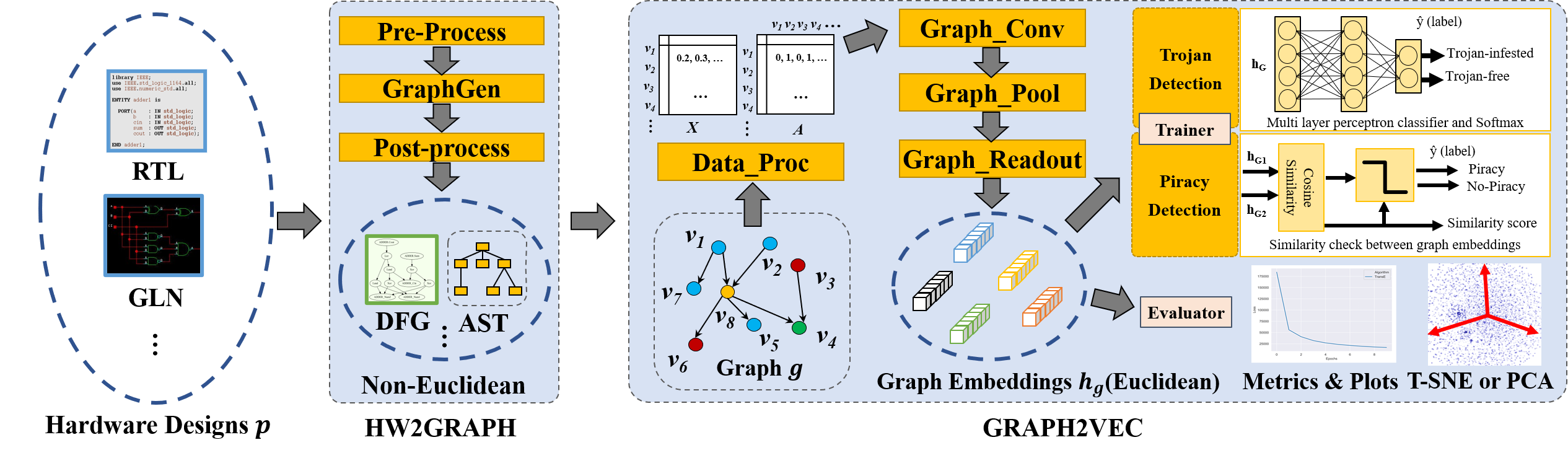}
    \caption{The overall architecture of \textsc{hw2vec}. Beginning with hardware design objects (RTL or GLN), the \textsc{HW2GRAPH} leverages \textsc{PRE\_PROC}, \textsc{GRAPH\_GEN}, and \textsc{POST\_PROC} to extract graph representations from hardware designs in the form of node embedding matrix ($\mathbf{X}$) and adjacency matrix ($\mathbf{A}$). These graphs are then passed to \textsc{GRAPH2VEC} to acquire the graph embeddings for graph learning tasks of hardware security.} 
    \label{fig:archi}
    \vspace{-1.0em}
\end{figure*}

As Figure~\ref{fig:archi} shows, \textsc{HW2VEC} contains \textsc{HW2GRAPH} and \textsc{GRAPH2VEC}. 
During the IC design flow, a hardware design can have various levels of abstraction such as High-Level Synthesis (HLS), RTL, GLN, and GDSII, each of which are fundamentally non-Euclidean data.
Overall, in \textsc{HW2VEC}, a hardware design $p$ is first turned into a graph $g$ by \textsc{HW2GRAPH}, which defines the pairwise relationships between objects that preserve the structural information. 
Then, \textsc{GRAPH2VEC} consumes $g$ and produces the Euclidean representation $h_g$ for learning.

\subsection{\textsc{HW2GRAPH: from hardware design to graph}}
\label{sec:hw2graph}
The first step is to convert each textual hardware design code $p$ into a graph $g$.
\textsc{HW2GRAPH} supports the automatic conversion of raw hardware code into various graph formats such as Abstract Syntax Tree (AST) or Data-Flow Graph (DFG).
AST captures the syntactic structure of hardware code while DFG indicates the relationships and dependencies between the signals and gives a higher-level expression of the code's computational structure. 
\textsc{HW2GRAPH} consists of three primary modules: \textit{pre-processing}, \textit{graph generation engine}, and \textit{post-processing}.
\subsubsection{Pre-processing (\textsc{PRE\_PROC})}
In this module, we have several automatic scripts for pre-processing a raw hardware code $p$.
As a hardware design can contain several modules stored in separate files, the first step is to combine them into a single file (i.e., flattening). 
Next, to automatically locate the ``entry point'' top module of $p$, the script scans the flattened code for the keyword ``module'' and extracts the module names and the number of repetitions in $p$.
Then, the script analyzes the list of discovered module names and takes the one that appears only once, which means the module is not instantiated by any other module, as the top module.
Here, we denote the pre-processed hardware design code as $p^\prime$.
\subsubsection{Graph Generation Engine (\textsc{GRAPH\_GEN})}
We integrate PyVerilog~\cite{takamaeda2015pyverilog}, a hardware design toolkit for parsing the Verilog code, into this module.
The pre-processed code $p^\prime$ is first converted by a lexical analyzer, YACC (Yet Another Compiler-Compiler), into a corresponding parse tree. 
Then, we recursively iterate through each node in the parse tree with Depth-First Search (DFS). 
At each recursive step, we determine whether to construct a collection of name/value pairs, an ordered list of values, or a single name/value pair based on the token names used in Verilog AST.
To acquire DFG, the AST is further processed by the data flow analyzer to create a signal DFG for each signal in the circuit such that the signal is the root node.
Lastly, we merge all the signal DFGs.
The resulting graph, either DFG or AST, is denoted as $g=(V, E)$.
The AST is a tree type of graph in which the nodes $V$ can be operators (mathematical, gates, loop, conditional, etc.), signals, or attributes of signals. The edges $E$ indicate the relation between nodes.
The DFG shows data dependency where each node in $V$ represents signals, constant values, and operations such as xor, and, concatenation, branch, or branch condition, etc. Each edge in $E$ stands for the data dependency relation between two nodes. 
Specifically,  for all $v_i, v_j$ pairs, the edge ${e_{ij}}$ belongs to $E$ (${e_{ij}}\in E$) if $v_i$ depends on $v_j$, or if $v_j$ is applied on $v_i$.
\subsubsection{Post-processsing (\textsc{POST\_PROC})} 
The output from \textit{Graph Generatifon Engine} is in JSON (JavaScript Object Notation) format. 
In this phase, we convert a JSON-formatted graph into a NetworkX graph object. 
NetworkX is an efficient, scalable, and highly portable framework for graph analysis. 
Several popular geometric representation learning libraries (PyTorch-Geometric and Deep Graph Library) take this format of graphs as the primary data structure in their pipelines.
\subsection{\textsc{Graph2Vec: from graph to graph embedding}}
\label{sec:graph2vec}

Once \textsc{hw2graph} has converted a hardware design into a graph $g$, we begin to process $g$ with the modules in \textsc{graph2vec}, including \textit{Dataset Processor}, \textit{Trainer}, and \textit{Evaluator} to acquire the graph embedding $h_g$.

\subsubsection{Dataset Processor} This module handles the low-level parsing tasks such as caching the data on disk to optimize the tasks that involve repetitive model testing, performing train-test split, finding the unique set of node labels among all the graph data instances.
One important task of the \textit{dataset processor} is to convert a graph $g=(V, E)$ into the tensor-like inputs $\mathbf{X}$ and $\mathbf{A}$ where $\mathbf{X}$ represents the node embeddings in matrix form and $\mathbf{A}$ stands for the adjacency information of $g$. 
The conversion between $E$ and $\mathbf{A}$ is straightforward. 
To acquire $\mathbf{X}$, \textit{Dataset Processor} performs a \textit{normalization} process and assigns each of the nodes a label that indicates its type, which may vary for different kinds of graphs (AST or DFG). 
Each node gets converted to an initial vectorized representation using one-hot encoding based on its type label.

\subsubsection{Graph Embedding Model}
In this module, we break down the graph learning pipeline into multiple network components, including graph convolution layers (\textit{GRAPH\_CONV}), graph pooling layers (\textit{GRAPH\_POOL}), and graph readout operations (\textit{GRAPH\_READOUT}). 

In \textsc{HW2VEC}, the \textit{GRAPH\_CONV} is inspired by the Spatial Graph Convolution Neural Network (SGCN), which defines the convolution operation based on a node's spatial relations. 
In literature, this phase is also referred to as \textit{message propagation phase} which involves two sub-functions: \textbf{AGGREGATE} and \textbf{COMBINE} functions. 
Each input graph $g=(V, E)$ is initialized in the form of node embeddings and adjacency information ($\mathbf{X}^{(0)}$ and $\mathbf{A}$).
For each $k$-th iteration, the process updates the node embeddings $\mathbf{X}^{(k)}$ using each node representation $h_v^{(k-1)}$ in $\mathbf{X}^{(k-1)}$, given by,
\begin{equation}
    a_v^{(k)} = \textbf{AGGREGATE}^{(k)}(\{h_u^{(k-1)}: u \in N(v)\})
\end{equation}
\begin{equation}
    h_v^{(k)} = \textbf{COMBINE}^{(k)}(h_v^{(k-1)}, a_v^{(k)} )
\end{equation}
where $h_v^{(k)} \in R^{C^k}$ denotes the feature vector after $k$ iterations for the $v$-th node and $N(v)$ returns the neighboring nodes of $v$-th node. 
Essentially, the \textbf{AGGREGATE} collects the features of the neighboring nodes to extract an aggregated feature vector $a_v^{(k)}$ for the layer k, and the \textbf{COMBINE} combines the previous node feature $h_v^{(k-1)}$ with  $a_v^{(k)}$ to output next feature vector $h_v^{(k)}$. 
This message propagation is carried out for a pre-determined number of layers $k$.
We denote the final propagation node embedding $\mathbf{X}^{(k)}$ as $\mathbf{X}^{prop}$. 

Next, in \textit{GRAPH\_POOL}, the node embedding $\mathbf{X}^{prop}$ is further processed with an attention-based graph pooling layer.
As indicated from \cite{lee2019self, ying2018hierarchical}, the integration of a graph pooling layer allows the model to operate on the hierarchical representations of a graph, and hence can better perform the graph classification task. 
Besides, such an attention-based pooling layer allows the model to focus on a local part of the graph and is considered as a part of a unified computational block of a GNN pipeline~\cite{knyazev2019understanding}.
In this layer, we perform \textit{top-k filtering} on nodes according to the scoring results, as follows: 
\begin{equation}
\mathbf{\alpha} = \textsc{SCORE}(\mathbf{X}^{prop}, \mathbf{A})
\end{equation}
\begin{equation}
    \mathbf{P} = \mathrm{top}_k(\mathbf{\alpha}) 
\end{equation}
where $\mathbf{\alpha}$ stands for the coefficients predicted by the graph pooling layer for nodes. $\mathbf{P}$ represents the indices of the pooled nodes, which are selected from the top $k$ of the nodes ranked according to $\alpha$. 
The number $k$ used in \textit{top-k filtering} is calculated by a pre-defined pooling ratio, $pr$ using $k = pr \times |V|$, where we consider only a constant fraction $pr$ of the embeddings of the nodes of the DFG to be relevant (i.e., 0.5). One example of the scoring function is to utilize a separate trainable GNN layer to produce the scores so that the scoring method considers both node features and topological characteristics~\cite{lee2019self}.
We denote the node embeddings and edge adjacency information after pooling by $\mathbf{X}^{pool}$ and $\mathbf{A}^{pool}$ which are calculated as follows:
\begin{equation}
\mathbf{X}^{pool}= (\mathbf{X}^{prop} \odot\mathrm{tanh}(\mathbf{\alpha}))_{\mathbf{P}} 
\end{equation}
\begin{equation}
\mathbf{A}^{pool} = {\mathbf{A}^{prop}}_{(\mathbf{P},\mathbf{P})} \\
\end{equation}
where $\odot$ represents an element-wise multiplication, $()_{\mathbf{P}}$ refers to the operation that extracts a subset of nodes based on $P$, and  $()_{(\mathbf{P},\mathbf{P})}$ refers to the information of the adjacency matrix between the nodes in this subset.

Lastly, in \textit{GRAPH\_READOUT}, the overall graph-level feature extraction is carried out by either summing up or averaging up the node features $\mathbf{X}^{pool}$. We denote the graph embedding for each graph $g$ as $h^{(k)}_g$, computed as follows: 
\begin{equation}
    h^{(k)}_g = \textit{GRAPH\_READOUT}(\{h_v^{(k)}: v \in V\})
\end{equation}
We use the graph embedding $h^{(k)}_g$ to model the behavior of circuits (use $h_g$ for simplicity). After this, the fixed-length embeddings of hardware designs then become compatible with ML algorithms.

In practice, these network components can be combined in various ways depending on the type of the tasks (node-level task, graph-level task) or the complexity of the tasks (simple or complex network architecture). 
In \textsc{GRAPH2VEC}, one default option is to use one or multiple \textit{GRAPH\_CONV}, followed by a \textit{GRAPH\_POOL} and a \textit{GRAPH\_READOUT}.
Besides, in conjunction with Multi-Layer Perceptron (MLP) or other ML layers, this architecture can transform the graph data into a form that we can use in calculating the loss for learning. 
In \textsc{GRAPH2VEC}, we reserve the flexibility for customization, so users may also choose to combine these components in a way that is effective for their tasks.

\subsubsection{Trainer and Evaluator} \label{sec:evaluator}
The \textit{Trainer} module takes training datasets, validating datasets, and a set of hyperparameter configurations to train a GNN model. 
\textit{HW2VEC} currently supports two types of \textit{Trainer}, \textit{graph-trainer} and \textit{graph-pair-trainer}. 
To be more specific, \textit{graph-trainer} uses \textsc{GRAPH2VEC}'s model to perform graph classification learning and evaluation while \textit{graph-pair-trainer} considers pairs of graphs, calculates their similarities, and ultimately performs the graph similarity learning and evaluation. 
Some low-level tasks are also handled by \textit{Trainer} module, such as caching the best model weights evaluated from the validation set to the disk space or performing mini-step testing.
Once the training is finished, the \textit{Evaluator} module plots the training loss and commonly used metrics in ML-based hardware security applications. To facilitate the analysis of the results, \textsc{HW2VEC} also provides utilities to visualize the embeddings of hardware designs with t-SNE based dimensionality reduction~\cite{van2008visualizing}. 
Besides, \textsc{HW2VEC} provides multiple exporting functionalities so that the learned embeddings can be presented in standardized formats, and users can also choose other third-party tools such as \textit{Embedding Projector}~\cite{smilkov2016embedding} to analyze the embeddings.
\section{HW2VEC Use-cases}
\label{sec:use-case}
In this section, we describe \textsc{HW2VEC} use-cases.
First, Section~\ref{sec:use-case-1} exhibits a fundamental use-case in which a hardware design $p$ is converted into a graph $g$ and then into a fixed-length embedding $h_g$. 
Next, the use-cases of \textsc{HW2VEC} for two hardware security applications (detecting hardware Trojan and hardware IP piracy) are described in Section~\ref{sec:use-case-2} and Section~\ref{sec:use-case-3}, respectively.
\subsection{Use-case 1: Converting a Hardware Design to a Graph Embedding}

\label{sec:use-case-1}
The first use-case demonstrates the transformation of a hardware design $p$ into a graph $g$ and then into an embedding $h_g$.
As Algorithm~\ref{alg:use-case-1} shows, \textsc{HW2GRAPH} uses \textit{preprocessing} (\textsc{PRE\_PROC}), \textit{graph generation} (\textsc{GRAPH\_GEN}) and \textit{post-processing} (\textsc{POST\_PROC}) modules which are detailed in Section~\ref{sec:hw2graph} to convert each hardware design into the corresponding graph.
The $g$ is fed to \textsc{GRAPH2VEC} with the uses of \textit{Data Processing} (\textsc{DATA\_PROC}) to generate $X$ and $A$.
Then, $X$ and $A$ are processed through \textit{GRAPH\_CONV}, \textit{GRAPH\_POOL}, and \textit{GRAPH\_READOUT} to generate the graph embedding $h_g$. 
This resulting $h_g$ can be further inspected with the utilities of \textit{Evaluator} (see Section~\ref{sec:evaluator}).
\begin{algorithm}[h]
    \SetAlgoLined
    \DontPrintSemicolon
    \textbf{Input:} A hardware design program $p$.\;
    \textbf{Output:} A graph embedding $h_p$ for $p$.\;
    \SetKwProg{Fn}{def}{:}{}
    \SetKwFunction{Fhwgraph}{\textit{HW2GRAPH}}
    \SetKwFunction{Fgraphvec}{\textit{GRAPH2VEC}}
    \Fn{\Fhwgraph{$p$}}{
        $p' \gets$ \textsc{Pre\_Proc}($p$);\;
        $g \gets$ \textsc{Graph\_Gen}($p'$);\;
        $g\prime \gets$ \textsc{Post\_Proc}($g$);\;
        \KwRet $g\prime$;\;
    }
    \Fn{\Fgraphvec{$g$}}{
        $X, A \gets$ \textsc{Data\_Proc}($g$)\;
        $X^{prop}, A^{prop} \gets$ \textit{GRAPH\_CONV}($X, A$)\;
        $X^{pool}, A^{pool} \gets$ \textit{GRAPH\_POOL}($X^{prop}, A^{prop} $)\;
        $h_{g} \gets$ \textit{GRAPH\_READOUT}($X^{pool}$)\;
        \KwRet $h_{g}$\;
    }
    $g \gets$ \textit{\texttt{HW2GRAPH}}($p$);\;
    $h_{g} \gets$ \textit{\texttt{GRAPH2VEC}}($g$);\;
    \caption{Use-case - \textsc{HW2VEC}}
    \label{alg:use-case-1}
\end{algorithm}
In \textsc{HW2VEC}, we provide Algorithm~\ref{alg:use-case-1}'s implementation in \texttt{use\_case\_1.py} of our repository. 

\subsection{Use-case 2: Hardware Trojan Detection}
\label{sec:use-case-2}

In this use-case, we demonstrate how to use \textsc{HW2VEC} to detect HT, which has been a major hardware security challenge for many years.
An HT is an intentional, malicious modification of a circuit by an attacker~\cite{rostami2013hardware}.
The capability of detection at an early stage (particularly at RTL level) is crucial as removing HTs at later stages could be very expensive.
The majority of existing solutions rely on a golden HT-free reference or cannot generalize detection to previously unseen HTs.
\cite{gnn4tj2021} proposes a GNN-based approach to model the circuit's behavior and identify the presence of HTs.

\begin{algorithm}[h]
    \SetAlgoLined
    \DontPrintSemicolon
    \textbf{Input:} A hardware design program $p$.\;
    \textbf{Output:} A label indicating whether $p$ contains Hardware Trojan.\;
    \SetKwProg{Fn}{def}{:}{}
    \SetKwFunction{Fsgvec}{\textit{use\_case\_2}}
    \Fn{\Fsgvec{$p$}}{
        $g \gets$ \textsc{HW2GRAPH}($p$);\;
        $h_g \gets$ \textsc{GRAPH2VEC}($g$);\;
        $\hat{y} \gets$ \textsc{MLP}($h_g$);\;
        \uIf{$\hat{y}[0] > \hat{y}[1] $}{
        \KwRet $\textsc{Trojan}$;\;}
        \uElse{
        \KwRet $\textsc{Non\_Trojan}$;\;}
    }
    $\hat{Y} \gets$ \textit{\texttt{use\_case\_2}}($p$);\;
    \caption{Use-case - Hardware Trojan Detection}
    \label{alg:use-case-2}
\end{algorithm}

To realize~\cite{gnn4tj2021} in \textsc{HW2VEC}, we first use \textsc{HW2GRAPH} to convert each hardware design $p$ into a graph $g$.
Then, we transform each $g$ to a graph embedding $h_g$.
Lastly, $h_g$ is used to make a prediction $\hat{y}$ with an MLP layer. 
To train the model, the cross-entropy loss $L$ is calculated collectively for all the graphs in the training set (see Equation~\ref{loss:cross}).
\begin{equation}L = H(Y, \hat{Y}) = \sum_{i} y_i * log_e(\hat{y_i}),\label{loss:cross}
\end{equation}
where $H$ is the loss function.
$Y$ stands for the set of ground-truth labels (either \textsc{Trojan} or \textsc{Non\_Trojan}) and $\hat{Y}$ represents the corresponding set of predictions.
Once trained by minimizing $L$, we use the model and Algorithm~\ref{alg:use-case-2} to perform HT detection (can also be done with a pre-trained model).
In practice, we provide an implementation in \texttt{use\_case\_2.py} in our repository. 

\subsection{Use-case 3: Hardware IP Piracy Detection}
\label{sec:use-case-3}
This use-case demonstrates how to leverage \textsc{HW2VEC} to confront another major hardware security challenge -- determining whether one of the two hardware designs is stolen from the other or not.
The IC supply chain has been so globalized that it exposes the IP providers to theft and illegal IP redistribution.
One state-of-the-art countermeasure embeds the signatures of IP owners on hardware designs (i.e., watermarking or fingerprinting), but it causes additional hardware overhead during the manufacturing.
Therefore,~\cite{gnn4ip2021} addresses IP piracy by assessing the similarities between hardware designs with a GNN-based approach.
Their approach models the behavior of a hardware design (in RTL or GLN) in graph representations.
\begin{algorithm}[h]
    \SetAlgoLined
    \DontPrintSemicolon
    \textbf{Input:} A pair of hardware design programs $p_1, p_2$.\;
    \textbf{Output:} A label indicating whether $p_1, p_2$ is piracy.\;
    \SetKwProg{Fn}{def}{:}{}
    \SetKwFunction{Fsgvec}{\textit{use\_case\_3}}

    \Fn{\Fsgvec{$p_1$, $p_2$}}{
        $g_1, g_2 \gets$ \textsc{HW2GRAPH}($p_1$), \textsc{HW2GRAPH}($p_2$);\;
        $h_{g_1}, h_{g_2} \gets$ \textsc{GRAPH2VEC}($g_1$), \textsc{GRAPH2VEC}($g_2$);\;
        $\hat{y} \gets$ \textsc{Cosine\_Sim}($h_{g_1}, h_{g_2}$);\;
        \uIf{$\hat{y} > \delta $}{
        \KwRet $\textsc{Piracy}$;\;}
        \uElse{
        \KwRet $\textsc{Non-Piracy}$;\;}
    }
    $\hat{Y} \gets$ \textit{\texttt{use\_case\_3}}($p_1$, $p_2$);\;
    \caption{Use-case - Hardware IP Piracy Detection}
    \label{alg:use-case-3}
\end{algorithm}

To implement~\cite{gnn4ip2021}, the GNN model has to be trained with a graph-pair classification trainer in \textsc{GRAPH2VEC}.
The first step is to use \textsc{HW2GRAPH} to convert a pair of circuit designs $p_1$, $p_2$ into a pair of graphs $g_1$, $g_2$.
Then, \textsc{GRAPH2VEC} transforms both $g_1$ and $g_2$ into graph embeddings $h_{g_1}$, $h_{g_2}$.
To train this GNN model for assessing the similarity of $h_{g_1}$ and $h_{g_2}$, a cosine similarity is computed as the final prediction of piracy, denoted as $\hat{y} \in [-1, 1]$. 
The loss between a prediction $\hat{y}$ and a ground-truth label $y$ is calculated as Equation~\ref{loss:cos_loss} shows. 
Lastly, the final loss $L$ is computed collectively with a loss function $H$ for all the graphs in the training set (see Equation~\ref{loss:sim_loss}).
\begin{equation}
\label{loss:cos_loss}
    G(y, \hat{y}) = \left \{
    \begin{array}{ll}
         1 - \hat{y}, & \texttt{if } y = 1\\
         \textsc{max}(0, \hat{y}-\textsc{margin}) & \texttt{if } y = -1
    \end{array}
    \right.
\end{equation} 
\begin{equation}
\label{loss:sim_loss}
    L = H(Y, \hat{Y}) = \sum_{i} G(y_i, \hat{y_i}),
\end{equation}
where $Y$ stands for the set of ground-truth labels (either \textsc{Piracy} or \textsc{Non\_Piracy}) and $\hat{Y}$ represents the corresponding set of predictions.
The \textsc{margin} is a constant to prevent the learned embedding from becoming distorted (always set to 0.5 in~\cite{gnn4ip2021}).
Once trained, we use this model and Algorithm~\ref{alg:use-case-3} with $\delta$, which is a decision boundary used for making final judgment, to detect piracy.
In practice, we provide the implementation of Algorithm~\ref{alg:use-case-3} in \texttt{use\_case\_3.py}.
\section{Experimental Results}
\label{sec:exp}
In this section, we evaluate the \textsc{HW2VEC} through various experiments using the use-case implementations described earlier.

\subsection{Dataset Preparation}
\label{sec:dataset}
For evaluation, we prepare one RTL dataset for HT detection (\textit{TJ-RTL}) and both RTL and GLN datasets (\textit{IP-RTL} and \textit{IP-GLN}) for IP piracy detection. 

\subsubsection{The TJ-RTL dataset} 
We construct the \textit{TJ-RTL} dataset by gathering the hardware designs with or without HT from the Trust-Hub.org benchmark~\cite{tehranipoor2016trusthub}.
From Trust-Hub, we collect three base circuits, \texttt{AES}, \texttt{PIC}, and \texttt{RS232}, and insert 34 varied types of HTs into them. 
We also include these HTs as standalone instances to the \textit{TJ-RTL} dataset. 
Furthermore, we insert these standalone HTs into two other circuits (\texttt{DES} and \texttt{RC5}) and include the resulting circuits to expand the \textit{TJ-RTL} dataset.
Among the five base circuits, \texttt{AES}, \texttt{DES}, and \texttt{RC5} are cryptographic cores that encrypt the input plaintext into the ciphertext based on a secret key.
For these circuits, the inserted HTs can leak sensitive information (i.e., secret key) via side-channels such as power and RF radiation or degrade the performance of their host circuits by increasing the power consumption and draining the power supply.
\texttt{RS232} is an implementation of the UART communication channel, while the HT attacks on \texttt{RS232} can affect the functionality of either transmitter or receiver or can interrupt/disable the communication between them. 
The \texttt{PIC16F84} is a well-known Power Integrated Circuit (PIC) microcontroller, and the HTs for PIC fiddle with its functionality and manipulate the program counter register.
Lastly, we create the graph datasets, \textit{DFG-TJ-RTL} and \textit{AST-TJ-RTL}, in which each graph instance is annotated with a \textsc{Trojan} or \textsc{Non\_Trojan} label. 

\subsubsection{The IP-RTL and IP-GNL datasets}
To construct the datasets for evaluating piracy detection, we gather RTL and GLN of hardware designs in Verilog format. 
The RTL dataset includes common hardware designs such as single-cycle and pipeline implementation of MIPS processor which are derived from available open-source hardware design in the internet or designed by a group of in-house designers who are given the same specification to design a hardware in Verilog. The GLN dataset includes ISCAS'85 benchmark~\cite{hansen1999unveiling} which includes 7 different hardware designs (\texttt{c432}, \texttt{c499}, \texttt{c880}, \texttt{c1355}, \texttt{c1908}, \texttt{c6288}, \texttt{c7552}) and their obfuscated instances derived from TrustHub. 
Obfuscation complicates the circuit and confuses reverse engineering but does not change the behavior of the circuit.
Our collection comprises 50 distinct circuit designs and several hardware instances for each circuit design that sums up 143 GLN and 390 RTL codes. 
We form a graph-pair dataset of 19,094 similar pairs and 66,631 different pairs, dedicate 20\% of these 85,725 pairs for testing and the rest for training.
This dataset comprises of pairs of hardware designs, labelled as \textsc{piracy} (positive) or \textsc{no-piracy} (negative).

\subsection{\textsc{HW2VEC} Evaluation: Hardware Trojan Detection}
\label{sec:eval_tj}
Here, we evaluate the capability of \textsc{HW2VEC} in identifying the existence of HTs from hardware designs. 
We leverage the implementation mentioned in Section~\ref{sec:use-case-2}.
As for hyperparameters, we follow the best setting used in~\cite{gnn4tj2021} which is stored as a preset in a \textsc{YAML} configuration file. 
For performance metrics, we count the True Positive ($TP$), False Negative ($FN$) and False Positive ($FP$) for deriving Precision $P = TP/(TP+FP)$ and Recall $R = TP/(TP+FN)$. 
$R$ manifests the percentage of HT-infested samples that the model can identify.
As the number of HT-free samples incorrectly classified as HT is also critical,
we compute $P$ that indicates what percentage of the samples that model classifies as HT-infested actually contains HT. 
$F_1$ score is the weighted average of precision and recall that better presents performance, calculated as $F_1 = 2 \times P \times R/(P+R)$.

To demonstrate whether the learned model can generalize the knowledge to handle the unknown or unseen circuits, we perform a variant \textit{leave-one-out} cross-validation to experiment.
We perform a train-test split on the \textit{TJ-RTL} dataset by leaving one base circuit benchmark in the testing set and use the remaining circuits to train the model.
We repeat this process for each base circuit and average the metrics we acquire from evaluating each testing set. 
The result is presented in Table~\ref{tab:htdetection}, indicating that \textsc{HW2VEC} can reproduce comparable results to~\cite{gnn4tj2021} in terms of $F_1$ score (0.926 versus 0.940) if we use DFG as the graph representation.
The difference in performance can be due to the use of different datasets. 
When using AST as the graph representation for detecting HT, \textsc{HW2VEC} performs worse in terms of $F_1$ score, indicating that DFG is a better graph representation because it captures the data flow information instead of simply the syntactic information of a hardware design code.
All in all, these results demonstrate that our \textsc{HW2VEC} can be leveraged for studying HT detection at design phases.

\begin{table}[!ht]
    \centering
    \begin{tabular}{c c c c c c c}
        \hline
        Method & Graph & Dataset & Precision & Recall & F1\\\hline
        \textsc{HW2VEC} & \textbf{DFG} & RTL & 0.87334 & 0.98572 & 0.92596 \\
        \textsc{HW2VEC} & \textbf{AST} & RTL & 0.90288 & 0.8 & 0.8453\\
        \cite{gnn4tj2021} & \textbf{DFG} & RTL & \textbf{0.923} & \textbf{0.966} & \textbf{0.940} \\
        \hline
    \end{tabular}
    \caption{The performance of HT detection using \textsc{HW2VEC}.}
    \label{tab:htdetection}
\end{table}

\subsection{\textsc{HW2VEC} Evaluation: Hardware IP Piracy Detection}
\label{sec:eval_ip}
Besides the capability of HT detection, we also evaluate the power of \textsc{HW2VEC} in detecting IP piracy. 
We leverage the usage example mentioned in Section~\ref{sec:use-case-3} which examines the cosine-similarity score $\hat{y}$ for each hardware design pair and produces the final prediction with the decision boundary. 
Using the \textit{IP-RTL} dataset and the \textit{IP-GNL} dataset (mentioned in Section~\ref{sec:dataset}), we generate graph-pair datasets by annotating the hardware designs that belong to the same hardware category as \textsc{Similar} and the ones that belong to different categories as \textsc{Dissimilar}. 
We perform a train-test split on the dataset so that 80\% of the pairs will be used to train the model.
We compute the accuracy of detecting hardware IP piracy, which expresses the correctly predicted sample ratio and calculates the $F_1$ score as the evaluating metrics.
We refer to~\cite{gnn4ip2021} for the selection of hyperparameters (stored in a YAML file).

\begin{figure}[!h]
    \centering
    \includegraphics[width=1.0\linewidth]{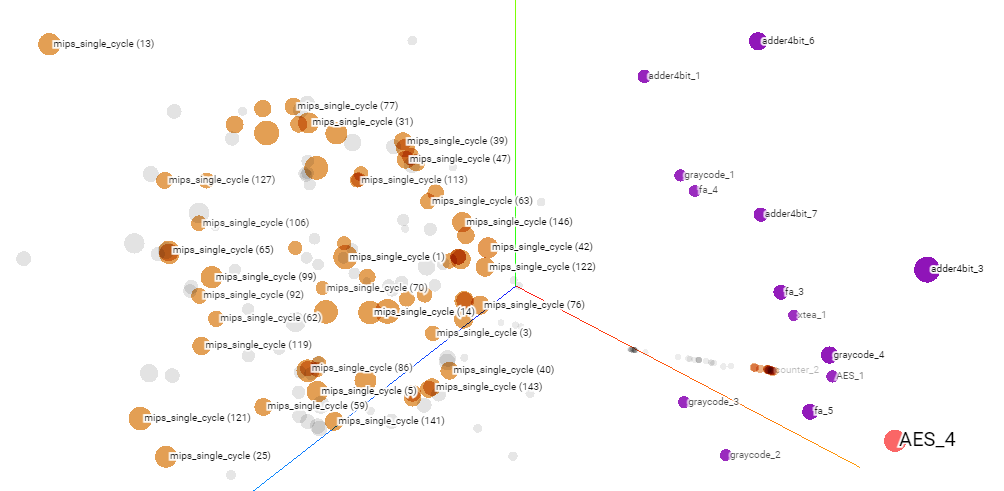}
    \caption{The embedding visualization with 3D t-SNE.}
    \label{fig:vis2}
\end{figure}

The result is presented in Table~\ref{tab:ipdetection}, indicating that \textsc{HW2VEC} can reproduce comparable results to~\cite{gnn4ip2021} in terms of piracy detection accuracy.
When using DFG as the graph representation, \textsc{HW2VEC} underperforms~\cite{gnn4ip2021} by 3\% at RTL level and outperforms~\cite{gnn4ip2021} by 4.2\% at GLN level.
Table~\ref{tab:ipdetection} also shows a similar observation with Section~\ref{sec:eval_tj} that using AST as the graph representation can lead to worse performance than using DFG.
Figure~\ref{fig:vis2} visualizes the graph embeddings that \textsc{HW2VEC} exports for every processed hardware design, allowing users to inspect the results manually.
For example, by inspecting Figure~\ref{fig:vis2}, we may find a clear separation between \texttt{mips\_single\_cycle} and \texttt{AES}. 
Certainly, \textsc{HW2VEC} can perform better with more fine-tuning processes.
However, the evaluation aims to demonstrate that \textsc{HW2VEC} can help practitioners study the problem of IP piracy at RTL and GLN levels.

\begin{table}[!ht]
    \centering
    \begin{tabular}{c c c c c c c}
        \hline
        Method & Graph & Dataset & Accuracy & F1 \\\hline
        \textsc{HW2VEC} & \textbf{DFG} & RTL & 0.9438 & 0.9277 \\
        \textsc{HW2VEC} & \textbf{DFG} & GLN & 0.9882 & 0.9652 \\
        \textsc{HW2VEC} & \textbf{AST} & RTL & 0.9358 & 0.9183 \\
        \cite{gnn4ip2021} & \textbf{DFG} & RTL & \textbf{0.9721} & --\\
        \cite{gnn4ip2021} & \textbf{DFG} & GLN & \textbf{0.9461} & --\\
        \hline
    \end{tabular}
    \caption{The results of detecting IP piracy with \textsc{HW2VEC}. }
    \label{tab:ipdetection}
\end{table}

\subsection{\textsc{HW2VEC} Evaluation: Timing}
To evaluate the time required for training and testing, we test the models on a server with NVIDIA TITAN-XP and NVIDIA GeForce GTX 1080 graphics cards.
Table \ref{tab:timing_model} indicates that the time taken by training and inference are both below 15 milliseconds, and the time taken by training is more than inference as it includes the time for performing back-propagation.
As \textsc{HW2VEC} aims to serve as a research tool, our users must evaluate their applications within a reasonable time duration. We believe that the time spent by the graph learning pipelines of \textsc{HW2VEC} should be acceptable for conducting research.
For practically deploying the models, the actual timing can depend on the computation power of hosting devices and the complexity of the models for the applications.
Suppose our users need an optimized performance for real-time applications. In that case, they can implement the models with performance-focused programming languages (C or C++) or ML frameworks (e.g., TensorFlow) using the best model settings found using \textsc{HW2VEC}.
As for specialized hardware that can accelerate the processing of GNNs, it is still an open challenge as indicated in~\cite{abadal2020computing}.

Table~\ref{tab:timing} indicates that the time that \textsc{HW2VEC} spends in converting the raw hardware code into ASTs is on average 1.98 seconds.
Although~\cite{han2019hardware} takes 1.37 seconds on average per hardware code, it requires domain knowledge to find a deterministic way to perform feature extraction.
For DFG extraction, \textsc{HW2VEC} takes on average 244.58 seconds per graph as it requires recursive traversals to construct the whole data flow.
In our datasets, \texttt{AES} and \texttt{DES} are relatively more complex, so \textsc{HW2VEC} takes 472.46 seconds on average processing them while the rest of the data instances take 16.70 seconds on average. 
Certainly, \textsc{HW2VEC} performs worse in DFG extraction, but manual feature engineering possibly requires a much longer time.
In design phases, even for an experienced hardware designer, it can take 6-9 months to prototype a complex hardware design~\cite{teel2017how} so the time taken by \textsc{HW2VEC} is acceptable and not slowing down the design process.
However, as the first open-source tool in the field, \textsc{HW2VEC} will keep evolving and embrace the contributions from the open-source community.

\begin{table}[h]
    \centering
    \begin{tabular}{c c c}
        \hline
         & \textit{TJ-RTL-AST} & \textit{IP-RTL-AST} \\ \hline
            training time & 10.5 (ms) & 13.5 (ms) \\ \hline
            testing time  & 6.8 (ms) & 12.4 (ms) \\ \hline
    \end{tabular}
\caption{The time profiling for training/inference.}
\label{tab:timing_model}
\end{table}

\begin{table}[h]
    \centering
    \begin{tabular}{c c c c}
        \hline
         & \textit{TJ-DFG-RTL} & \textit{IP-DFG-GLN} & \textit{TJ-AST-RTL} \\ \hline
        \# of node  & 7573.58 & 7616.16 & 971.01 \\ \hline
        \# of edge  & 8938.11 & 9495.97 & 970.01 \\ \hline
        Exec time & 244.58 (s) & 14.61 (s) & 1.98 (s) \\ \hline
    \end{tabular}
\caption{The graph extraction time profiling. For \textit{TJ-DFG-RTL}, the hardware \texttt{AES} and \texttt{DES} jointly take 472.46 seconds on average for DFG extraction while the rest of data instances take 16.7 seconds on average.}
\label{tab:timing}
\end{table}

\subsection{\textsc{HW2VEC} Applicability}
In Section~\ref{sec:eval_tj} and Section~\ref{sec:eval_ip}, we have discussed the performance of the GNN-based approach in resolving two hardware security problems: hardware Trojan detection and IP piracy detection.
In Section~\ref{sec:eval_tj}, our evaluation shows that \textsc{HW2VEC} can successfully be leveraged to perform HT detection on hardware designs, particularly on the unseen ones, without the assistance of golden HT-free reference.
The capability to model hardware behaviors can be attributed to using a natural representation of the hardware design (e.g., DFG) and the use of the GNN-based method for capturing both the structural information and semantic information from the DFG and co-relating this information to the final HT labels.
Similarly, Section~\ref{sec:eval_ip} indicates that \textsc{HW2VEC} can be utilized to assess the similarities between circuits and thus can be a countermeasure for IP piracy. The use of graph representation for a hardware design and a Siamese GNN-based network architecture are the keys in~\cite{gnn4ip2021} to perform IP piracy detection at both RTL and GLN levels. 
For other hardware security applications, the flexible modules provided by \textsc{HW2VEC} (\textit{Trainer} and \textit{Evaluator}) can be adapted easily to different problem settings. 
For example, by adjusting the \textit{Trainer} to train the GNN models for node classification, \textsc{HW2VEC} can be adapted to localize the HT(s) or hardware bug(s) that exist in the hardware designs.
Also, the cached models provided by \textsc{HW2VEC} can be used in learning other new hardware design related tasks through the transfer of knowledge from a related task that has already been learned as the idea of \textit{Transfer Learning} suggests~\cite{torrey2010transfer}.
\section{Conclusion}
\label{sec:conclusion}
As technological advancements continue to grow, the fights between attackers and defenders will rise in complexity and severity.
To contribute to the hardware security research community, we propose \textsc{HW2VEC}: a graph learning tool for automating hardware security.
\textsc{HW2VEC} provides an automated pipeline for hardware security practitioners to extract graph representations from a hardware design in either RTL or GLN. Besides, the toolbox of \textsc{HW2VEC} allows users to realize their hardware security applications with flexibility. 
Our evaluation shows that \textsc{HW2VEC} can be leveraged and integrated for counteracting two critical hardware security threats: \textit{Hardware Trojan Detection} and \textit{IP Piracy Detection}.
Lastly, as discussed in this paper, we anticipate that \textsc{HW2VEC} can provide more straightforward access for both practitioners and researchers to apply graph learning approaches to hardware security applications.

\bibliographystyle{abbrv}
\bibliography{hw2vec.bib}

\begin{thebibliography}{10}

\bibitem{tehranipoor2016trusthub}
Trusthub.
\newblock {\em Available on-line: https://www.trust-hub.org}, 2016.

\bibitem{news2}
Special 301 report.
\newblock {\em the United States Trade Representative}, 2017.

\bibitem{abadal2020computing}
S.~Abadal, A.~Jain, R.~Guirado, J.~L{\'o}pez-Alonso, and E.~Alarc{\'o}n.
\newblock Computing graph neural networks: A survey from algorithms to
  accelerators.
\newblock {\em arXiv preprint arXiv:2010.00130}, 2020.

\bibitem{syrianRadar}
S.~Adee.
\newblock The hunt for the kill switch.
\newblock In {\em IEEE Spectrum}, 2008.

\bibitem{ashraf2018towards}
M.~AshrafiAmiri~et al.
\newblock Towards side channel secure cyber-physical systems.
\newblock In {\em Real-Time and Embedded Systems and Technologies}, 2018.

\bibitem{board2005defense}
D.~Board.
\newblock Defense science board (dsb) study on high performance microchip
  supply.
\newblock {\em URL www. acq. osd. mil/dsb/reports/ADA435563. pdf,[March 16,
  2015]}, 2005.

\bibitem{cai2018comprehensive}
H.~Cai, V.~W. Zheng, and K.~C.-C. Chang.
\newblock A comprehensive survey of graph embedding: Problems, techniques, and
  applications.
\newblock {\em IEEE Transactions on Knowledge and Data Engineering},
  30(9):1616--1637, 2018.

\bibitem{chen2020decoy}
J.~Chen~et al.
\newblock Decoy: Deflection-driven hls-based computation partitioning for
  obfuscating intellectual property.
\newblock In {\em Design Automation Conference (DAC)}, 2020.

\bibitem{htnet}
S.~Faezi, R.~Yasaei, and M.~Al~Faruque.
\newblock Htnet: Transfer learning for golden chip-free hardware trojan
  detection.
\newblock {\em IEEE/ACM Design Automation and Test in Europe Conference
  (DATE'21)}, 2021.

\bibitem{htm}
S.~Faezi~et al.
\newblock Brain-inspired golden chip free hardware trojan detection.
\newblock {\em IEEE Transaction on Information Forensics and Security (IEEE
  TIFS’21)}, 2021.

\bibitem{han2019hardware}
T.~Han, Y.~Wang, and P.~Liu.
\newblock Hardware trojans detection at register transfer level based on
  machine learning.
\newblock In {\em 2019 IEEE International Symposium on Circuits and Systems
  (ISCAS)}, pages 1--5. IEEE, 2019.

\bibitem{hansen1999unveiling}
M.~C. Hansen, H.~Yalcin, and J.~P. Hayes.
\newblock Unveiling the iscas-85 benchmarks: A case study in reverse
  engineering.
\newblock {\em IEEE Design \& Test of Computers}, 16(3):72--80, 1999.

\bibitem{hasegawa2018hardware}
K.~Hasegawa, Y.~Shi, and N.~Togawa.
\newblock Hardware trojan detection utilizing machine learning approaches.
\newblock In {\em 2018 17th IEEE International Conference On Trust, Security
  And Privacy In Computing And Communications/12th IEEE International
  Conference On Big Data Science And Engineering (TrustCom/BigDataSE)}, pages
  1891--1896. IEEE, 2018.

\bibitem{herder2014physical}
C.~Herder, M.-D. Yu, F.~Koushanfar, and S.~Devadas.
\newblock Physical unclonable functions and applications: A tutorial.
\newblock {\em Proceedings of the IEEE}, 102(8):1126--1141, 2014.

\bibitem{hu2020overview}
W.~Hu, C.-H. Chang, A.~Sengupta, S.~Bhunia, R.~Kastner, and H.~Li.
\newblock An overview of hardware security and trust: Threats, countermeasures
  and design tools.
\newblock {\em IEEE Transactions on Computer-Aided Design of Integrated
  Circuits and Systems}, 2020.

\bibitem{huang2012parametric}
K.~Huang, J.~M. Carulli, and Y.~Makris.
\newblock Parametric counterfeit ic detection via support vector machines.
\newblock In {\em 2012 IEEE International Symposium on Defect and Fault
  Tolerance in VLSI and Nanotechnology Systems (DFT)}, pages 7--12. IEEE, 2012.

\bibitem{cadense2013jasper}
Jasper.
\newblock Jaspergold: Security path verification app.
\newblock 2014.

\bibitem{jose2008innovation}
S.~Jose.
\newblock Innovation is at risk as semiconductor equipment and materials.
\newblock {\em Semiconductor Equipment and Material Industry (SEMI)}, 2008.

\bibitem{kipf2016semi}
T.~N. Kipf and M.~Welling.
\newblock Semi-supervised classification with graph convolutional networks.
\newblock {\em arXiv preprint arXiv:1609.02907}, 2016.

\bibitem{knyazev2019understanding}
B.~Knyazev~et al.
\newblock Understanding attention and generalization in graph neural networks.
\newblock In {\em Advances in Neural Information Processing Systems (NeurIPS)},
  2019.

\bibitem{koushanfar2017active}
F.~Koushanfar.
\newblock Active hardware metering by finite state machine obfuscation.
\newblock In {\em Hardware Protection through Obfuscation}. 2017.

\bibitem{kulkarni2016svm}
A.~Kulkarni, Y.~Pino, and T.~Mohsenin.
\newblock Svm-based real-time hardware trojan detection for many-core platform.
\newblock In {\em 2016 17th International Symposium on Quality Electronic
  Design (ISQED)}, pages 362--367. IEEE, 2016.

\bibitem{lee2019self}
J.~Lee~et al.
\newblock Self-attention graph pooling.
\newblock {\em arXiv preprint arXiv:1904.08082}, 2019.

\bibitem{ma2020understanding}
Y.~Ma, Z.~He, W.~Li, L.~Zhang, and B.~Yu.
\newblock Understanding graphs in eda: From shallow to deep learning.
\newblock In {\em ISPD}, pages 119--126, 2020.

\bibitem{ma2019high}
Y.~Ma, H.~Ren, B.~Khailany, H.~Sikka, L.~Luo, K.~Natarajan, and B.~Yu.
\newblock High performance graph convolutional networks with applications in
  testability analysis.
\newblock In {\em Proceedings of the 56th Annual Design Automation Conference
  2019}, pages 1--6, 2019.

\bibitem{patnaik2018raise}
S.~Patnaik~et al.
\newblock Raise your game for split manufacturing: Restoring the true
  functionality through beol.
\newblock In {\em Design Automation Conference (DAC)}, 2018.

\bibitem{poudel2020flashmark}
P.~Poudel~et al.
\newblock Flashmark: watermarking of nor flash memories for counterfeit
  detection.
\newblock In {\em Design Automation Conference (DAC)}, 2020.

\bibitem{rahman2014ti}
M.~T. Rahman, K.~Xiao, D.~Forte, X.~Zhang, J.~Shi, and M.~Tehranipoor.
\newblock Ti-trng: Technology independent true random number generator.
\newblock In {\em 2014 51st ACM/EDAC/IEEE Design Automation Conference (DAC)},
  pages 1--6. IEEE, 2014.

\bibitem{rai2019hardware}
S.~Rai~et al.
\newblock Hardware watermarking using polymorphic inverter designs based on
  reconfigurable nanotechnologies.
\newblock In {\em 2019 IEEE Computer Society Annual Symposium on VLSI
  (ISVLSI)}, 2019.

\bibitem{camouflaging}
J.~Rajendran~et al.
\newblock Security analysis of integrated circuit camouflaging.
\newblock In {\em ACM conference on Computer \& communications security}, 2013.

\bibitem{rajendran2015detecting}
J.~Rajendran~et al.
\newblock Detecting malicious modifications of data in third-party intellectual
  property cores.
\newblock In {\em ACM/IEEE Design Automation Conference (DAC)}, 2015.

\bibitem{rajendran2016formal}
J.~Rajendran~et al.
\newblock Formal security verification of third party intellectual property
  cores for information leakage.
\newblock In {\em International Conference on VLSI Design and Embedded Systems
  (VLSID)}, 2016.

\bibitem{rostami2014primer}
M.~Rostami, F.~Koushanfar, and R.~Karri.
\newblock A primer on hardware security: Models, methods, and metrics.
\newblock {\em Proceedings of the IEEE}, 102(8):1283--1295, 2014.

\bibitem{rostami2013hardware}
M.~Rostami, F.~Koushanfar, J.~Rajendran, and R.~Karri.
\newblock Hardware security: Threat models and metrics.
\newblock In {\em 2013 IEEE/ACM International Conference on Computer-Aided
  Design (ICCAD)}, pages 819--823. IEEE, 2013.

\bibitem{shamsi2019ip}
K.~Shamsi, M.~Li, K.~Plaks, S.~Fazzari, D.~Z. Pan, and Y.~Jin.
\newblock Ip protection and supply chain security through logic obfuscation: A
  systematic overview.
\newblock {\em ACM Transactions on Design Automation of Electronic Systems
  (TODAES)}, 24(6):1--36, 2019.

\bibitem{smilkov2016embedding}
D.~Smilkov, N.~Thorat, C.~Nicholson, E.~Reif, F.~B. Vi{\'e}gas, and
  M.~Wattenberg.
\newblock Embedding projector: Interactive visualization and interpretation of
  embeddings.
\newblock {\em arXiv preprint arXiv:1611.05469}, 2016.

\bibitem{subramanyan2014formal}
P.~Subramanyan and D.~Arora.
\newblock Formal verification of taint-propagation security properties in a
  commercial soc design.
\newblock In {\em Design, Automation \& Test in Europe Conference (DATE)},
  2014.

\bibitem{Takamaeda:2015:ARC:Pyverilog}
S.~Takamaeda-Yamazaki.
\newblock Pyverilog: A python-based hardware design processing toolkit for
  verilog hdl.
\newblock In {\em Applied Reconfigurable Computing}, volume 9040 of {\em
  Lecture Notes in Computer Science}, pages 451--460. Springer International
  Publishing, Apr 2015.

\bibitem{takamaeda2015pyverilog}
S.~Takamaeda-Yamazaki.
\newblock Pyverilog: A python-based hardware design processing toolkit for
  verilog hdl.
\newblock In {\em International Symposium on Applied Reconfigurable Computing},
  2015.

\bibitem{tan2020challenges}
B.~Tan and R.~Karri.
\newblock Challenges and new directions for ai and hardware security.
\newblock In {\em 2020 IEEE 63rd International Midwest Symposium on Circuits
  and Systems (MWSCAS)}, pages 277--280. IEEE, 2020.

\bibitem{teel2017how}
J.~Teel.
\newblock How long does it take to develop a new product and get it to market?
\newblock Oct 2017.

\bibitem{torrey2010transfer}
L.~Torrey and J.~Shavlik.
\newblock Transfer learning.
\newblock In {\em Handbook of research on machine learning applications and
  trends: algorithms, methods, and techniques}, pages 242--264. IGI global,
  2010.

\bibitem{van2008visualizing}
L.~Van~der Maaten and G.~Hinton.
\newblock Visualizing data using t-sne.
\newblock {\em Journal of machine learning research}, 9(11), 2008.

\bibitem{waksman2013fanci}
A.~Waksman~et al.
\newblock Fanci: identification of stealthy malicious logic using boolean
  functional analysis.
\newblock In {\em ACM SIGSAC Conference on Computer and Communications
  Security}, 2013.

\bibitem{wu2020comprehensive}
Z.~Wu~et al.
\newblock A comprehensive survey on graph neural networks.
\newblock {\em IEEE Transactions on Neural Networks and Learning Systems},
  2020.

\bibitem{xiao2016hardware}
K.~Xiao, D.~Forte, Y.~Jin, R.~Karri, S.~Bhunia, and M.~Tehranipoor.
\newblock Hardware trojans: Lessons learned after one decade of research.
\newblock {\em ACM Transactions on Design Automation of Electronic Systems
  (TODAES)}, 22(1):1--23, 2016.

\bibitem{xie2017delay}
Y.~Xie~et al.
\newblock Delay locking: Security enhancement of logic locking against ic
  counterfeiting and overproduction.
\newblock In {\em Design Automation Conference (DAC)}, 2017.

\bibitem{gnn4ip2021}
R.~Yasaei, S.-Y. Yu, and M.~A.~A. Faruque.
\newblock Gnn4ip: Graph neural network for hardware intellectual property
  piracy detection.
\newblock In {\em Design, Automation \& Test in Europe Conference \& Exhibition
  (DATE)}. Ieee, 2021.

\bibitem{gnn4tj2021}
R.~Yasaei, S.-Y. Yu, and M.~A.~A. Faruque.
\newblock Gnn4tj: Graph neural networks for hardware trojan detection at
  register transfer level.
\newblock In {\em Design, Automation \& Test in Europe Conference \& Exhibition
  (DATE)}. Ieee, 2021.

\bibitem{yeh2012trends}
A.~Yeh.
\newblock Trends in the global ic design service market.
\newblock {\em DIGITIMES research}, 2012.

\bibitem{ying2018hierarchical}
R.~Ying, J.~You, C.~Morris, X.~Ren, W.~L. Hamilton, and J.~Leskovec.
\newblock Hierarchical graph representation learning with differentiable
  pooling.
\newblock {\em arXiv preprint arXiv:1806.08804}, 2018.

\bibitem{zhang2020grannite}
Y.~Zhang, H.~Ren, and B.~Khailany.
\newblock Grannite: Graph neural network inference for transferable power
  estimation.
\newblock In {\em 2020 57th ACM/IEEE Design Automation Conference (DAC)}, pages
  1--6. IEEE, 2020.

\bibitem{zhang2018analysis}
B.~Zhang~et al.
\newblock Analysis of security of split manufacturing using machine learning.
\newblock In {\em Design Automation Conference (DAC)}, 2018.

\bibitem{zhang2015veritrust}
J.~Zhang~et al.
\newblock Veritrust: Verification for hardware trust.
\newblock {\em IEEE Tran. on Computer-Aided Design of Integrated Circuits and
  Systems}, 2015.

\end{thebibliography}
\end{document}